\documentclass[letterpaper,pdftex,rmp,twocolumn,amsmath,amssymb,groupedaddress,floatfix]{revtex4}
\usepackage[scaled]{helvet}
\usepackage[utf8]{inputenc}
\usepackage[T1]{fontenc}
\usepackage{ae}
\usepackage{aecompl}
\usepackage{graphicx}
\usepackage{natbib}
\usepackage{color}
\usepackage[dvipsnames]{xcolor}
\usepackage{textcomp}
\usepackage{makecell}
\usepackage{mathptmx}
\usepackage{float}
\usepackage{setspace}
\usepackage[a]{esvect}
\usepackage{fixmath}
\usepackage[mathcal]{euscript}
\usepackage{scalerel}

\usepackage{caption}
\DeclareCaptionLabelSeparator{bar}{ \rule[-0.2em]{0.3ex}{1em} }
\usepackage[sf,bf,small,noindentafter]{titlesec}
\titlespacing*{\section}{0pt}{1em}{0em}
\titleformat{\subsection}[runin]{\bfseries}{}{}{}

\newcommand{\colorrule}[3]{
  \textcolor{#1}{\rule{#2}{#3}}
}
\definecolor{lightgray}{rgb}{0.66,0.66,0.55}
\definecolor{darkgray}{rgb}{0.331,0.331,0.276}
\definecolor{myblue}{HTML}{7FB0D1}
\definecolor{mygreen}{HTML}{A9D163}

\usepackage{url}

\usepackage[pdftex,breaklinks=true,colorlinks=true,citecolor=black,linkcolor=black,menucolor=black,urlcolor=myblue,pdfborder={1 0 0}]{hyperref}
\hypersetup{pdftitle={Maps of sparse Markov chains efficiently reveal community structure in network flows with memory},pdfauthor={C. Persson, L. Bohlin, D. Edler, and M. Rosvall (2016)}}
\bibpunct{}{}{,}{s}{,}{,}

\DeclareMathOperator*{\argmin}{\arg\,\min}

\usepackage{booktabs}

\newcommand{\mytoprule}{\specialrule{0.1em}{0em}{0em}}
\newcommand{\mybottomrule}{\specialrule{0.1em}{0em}{0em}} 
\newcommand{\mymidrule}{\specialrule{0.05em}{0em}{0em}}

\usepackage[activate={true,nocompatibility},final,tracking=true,kerning=true,spacing=true,factor=1100,stretch=10,shrink=10]{microtype}
\microtypecontext{spacing=nonfrench}

\begin{document}
\makeatletter
\renewcommand\@biblabel[1]{#1.}

\newsavebox\myboxA
\newsavebox\myboxB
\newlength\mylenA

\newcommand*\xoverline[2][0.75]{%
    \sbox{\myboxA}{$\m@th#2$}%
    \setbox\myboxB\null
    \ht\myboxB=\ht\myboxA%
    \dp\myboxB=\dp\myboxA%
    \wd\myboxB=#1\wd\myboxA
    \sbox\myboxB{$\m@th\overline{\copy\myboxB}$}
    \setlength\mylenA{\the\wd\myboxA}
    \addtolength\mylenA{-\the\wd\myboxB}%
    \ifdim\wd\myboxB<\wd\myboxA%
       \rlap{\hskip 0.5\mylenA\usebox\myboxB}{\usebox\myboxA}%
    \else
        \hskip -0.5\mylenA\rlap{\usebox\myboxA}{\hskip 0.5\mylenA\usebox\myboxB}%
    \fi}

\makeatother
    
\renewcommand{\figurename}{Figure}
\renewcommand{\thefigure}{\arabic{figure}}
\renewcommand{\tablename}{Table}
\renewcommand{\thetable}{\arabic{table}}
\renewcommand{\refname}{\large References}

\addtolength{\textheight}{1cm}
\addtolength{\textwidth}{1cm}
\addtolength{\hoffset}{-0.5cm}

\setlength{\belowcaptionskip}{1ex}
\setlength{\textfloatsep}{2ex}
\setlength{\dbltextfloatsep}{2ex}

\hyphenation{page-rank}

\newcommand*{\citen}[1]{%
  \begingroup
    \romannumeral-`\x 
    \setcitestyle{numbers}%
    \cite{#1}%
  \endgroup   
}

\title{Maps of sparse Markov chains efficiently reveal community structure\\ in network flows with memory}

\author{Christian Persson}
\author{Ludvig Bohlin}
\author{Daniel Edler}
\author{Martin Rosvall}%
\thanks{Corresponding author}
\email{martin.rosvall@umu.se}
\affiliation{%
Integrated Science Lab, Department of Physics, Ume{\aa} University, SE-901 87 Ume{\aa}, Sweden
}%

\begin{abstract}
\normalsize To better understand the flows of ideas or information through social and biological systems, researchers develop maps that reveal important patterns in network flows. In practice, network flow models have implied memoryless first-order Markov chains, but recently researchers have introduced higher-order Markov chain models with memory to capture patterns in multi-step pathways. Higher-order models are particularly important for effectively revealing actual, overlapping community structure, but higher-order Markov chain models suffer from the curse of dimensionality: their vast parameter spaces require exponentially increasing data to avoid overfitting and therefore make mapping inefficient already for moderate-sized systems. To overcome this problem, we introduce an efficient cross-validated mapping approach based on network flows modeled by sparse Markov chains. To illustrate our approach, we present a map of citation flows in science with research fields that overlap in multidisciplinary journals. Compared with currently used categories in science of science studies, the research fields form better units of analysis because the map more effectively captures how ideas flow through science.
\end{abstract}

\maketitle

\noindent For studying and better understanding interconnected social and biological systems, it is essential to simplify and highlight their flows of ideas, information, money, people, or goods with maps of network flows.\cite{brin1998anatomy,vespignani2012modelling} Good maps should compress the flows by downplaying noise and highlighting important regularities, such as modules in which flows persist for a long time, and they ultimately rely on effective models of network flows. In network science, researchers traditionally model network flows with random walks on the networks. While such a memoryless first-order Markov chain model is sufficient to capture the flow dynamics in some systems, recent studies have shown that a variety of integrated systems require higher-order Markov chain models with memory to capture important flow patterns,\cite{song2010limits,belik2011natural,pfitzner2013betweenness} such as high return flows essential for revealing actual, overlapping community structure.\cite{rosvall2014memory,salnikov2016using,Xu:2016ex} But the increasing complexity of fixed higher-order Markov chain models comes with three significant drawbacks: First, the models grow exponentially with Markov order and quickly become computationally inefficient. Second, the exponentially larger models require exponentially more data for statistically sound fits. Third, in practice there are too few models for good fits to the data; an mth-order model may underfit and an (m+1)th-order model overfit the data. Inevitably, the problems with fixed higher-order Markov chain models will transfer to methods that build on them, including compression and prediction, as well as mapping network flows with memory.

To overcome these problems of fixed higher-order Markov chain models and design algorithms for better compression and prediction of sequence data, such as DNA or text, researchers have introduced variable-order Markov chain models. They allow the amount of memory to vary depending on the state of the flow, that is, where a flow entity is and came from.\cite{rissanen1986complexity,Buhlmann:1999iw,begleiter2004prediction,BORGES:2005jv,Chierichetti:2012ge,Singer:2014dk} While some states lack data for anything but a memoryless model, other states may be supported with more data and also require more memory for a good fit. Where fixed-order Markov chain models completely fail, variable-order Markov chain models can succeed thanks to their structural richness. But these models are designed for sequences with the small alphabets of DNA or text, and not for networks with thousands of nodes.  Recently, researchers introduced so-called higher-order networks,\cite{Xu:2016ex} but they are general-purpose representations for existing network methods. This leaves open the question of how to best integrate variable-order Markov chain models for efficiently mapping network flows with memory.

We design maps based on the most general class of variable-order Markov chain models, so-called sparse Markov chains.\cite{Jaaskinen:2014he} To select the best model for mapping a given set of network flow data, we first iteratively lump states based on minimum entropy rate loss down to a first-order Markov chain model. This procedure gives us an efficient description of the network flows for any number of states. To identify the optimal number of states for the map, we cluster sparse Markov chains with increasing number of states into modules with long flow persistence until we, according to ten-fold cross-validation, obtain the best map for a modular description of the data. We first describe these steps in detail below, and for illustration then apply the method to citation flows. Compared with an established journal classification by Web of Science, we find that a map of network flows modeled with sparse Markov chains almost doubles the module flow persistence---thereby providing a more accurate modular description of the citation flows in science.


\section*{Results}
\subsection*{Modeling networks flows with sparse Markov chains.}
Conventional stochastic models of dynamics on networks assume a stationary first-order Markov chain. That is, for a random walker that steps between the concrete objects that flow entities can visit, the \emph{physical nodes} $i \in \chi = \{1,\ldots,N\}$ over time $t$, and generates a sequence of random variables $X_1,\ldots,X_t$, the transition probabilities only depend on the previously visited node,
\begin{align}
\begin{split}
&P(X_t | X_{t-1}, X_{t-2}, \ldots ) = \\
&P(X_t | X_{t-1}).
\end{split}
\end{align}
For this time-homogeneous first-order Markov chain, the link weights $w_{ij}$ between physical nodes $i$ and $j$ normalized by the total weight of outgoing links $w_i = \sum_{j \in \chi}w_{ij}$ give the first-order transition probabilities
\begin{align}
P_{ij} = \frac{w_{ij}}{w_{i}} \mathrm{\ for\ } i,j \in \chi.\label{eq:firstordertransitionprob}
\end{align}

The straightforward generalization of first-order Markov chains is fixed higher-order Markov chains with memory of several previously visited physical nodes. They have a long history in, for example, computer science, statistics, and bioinformatics for compression or prediction, but have only recently been introduced to network science for epidemics, ranking, and community detection\cite{balcan2011phase,belik2011natural,poletto2013human,rosvall2014memory,peixoto2015modeling,salnikov2016using}. They were introduced because first-order Markov chains do not account for some important phenomena in network dynamics, such as high return flow that confines flow in smaller and more overlapping modules.\cite{rosvall2014memory} For a Markov chain of order $m$, the transition probabilities
\begin{align}
\begin{split}
&P(X_t | X_{t-1}, X_{t-2}, \ldots ) = \\
&P(X_t | X_{t-1}, \ldots, X_{t-m})
\label{eq:homarkov}
\end{split}
\end{align}
depend on the $m$ previously visited physical nodes. Any Markov chain of order $m$ can be represented as a first-order Markov chain with \emph{state nodes} $u \in \chi^m$ given by all possible subsequences $X_{t-m},\ldots,X_{t-1}$. In this way, the $m$th-order transition probabilities between physical nodes $i$ and $j$, $P_{ij}$, correspond to first-order transition probabilities between state nodes $u = X_{t-m},\ldots,X_{t-2},i$ and $v = X_{t-m+1},\ldots,X_{t-2},i,j$,
\begin{align}
P_{ij} = P_{uj} = P_{uv} = \frac{w_{uv}}{w_{u}} \mathrm{\ for\ } u,v \in \chi^m.\label{eq:higherordertransitionprob}
\end{align}

While higher-order Markov chains can capture important phenomena, model selection becomes problematic because the number of state nodes increases exponentially with Markov order and often prevents computation. Moreover, there are often insufficient data for the vast parameter space, yet there are very few members in the model class; a second-order Markov chain may underfit the data while a third-order Markov chain may overfit the data with no good model in between.\cite{Machler:2012hd} Therefore, researchers have introduced different classes of variable-length Markov chains.\cite{rissanen1983universal,Buhlmann:1999iw,Jaaskinen:2014he} Variable-length Markov chains lump together states with similar transition probabilities into single states such that the Markov chains will have memory of variable length. This gives a more compact description of the data and vastly increases the number of members in the model class such that it will be possible to fit models in-between fixed-length Markov chains. These properties make variable-length Markov chains highly efficient for compressing and predicting sequence data, and they have successfully been applied in bioinformatics and natural language modeling,\cite{Xiong:2016co} but only very recently in network science.\cite{Xu:2016ex}

Different classes of variable-length Markov chains impose different constraints on the state lumping, and the unconstrained and therefore most general class is called sparse Markov chains.\cite{Jaaskinen:2014he} With primed state nodes $u'$ and $v'$ for the fixed $m$th-order Markov chain model, its sparse Markov chain model is defined by a partition $S = \{s_1,\ldots,s_r\}$ of all state nodes $u' \in \chi^m$ into $r < |\chi^m|$ lumped state nodes $u' \in s_u \to u \subset \chi^m$ and the corresponding set of $r$ lumped transition probabilities $P_u$. For a sparse Markov chain given by partition $S$, the lumped link weights give the corresponding transition probabilities,
\begin{align}
P_{uv} = \frac{w_{uv}}{w_{u}} = \frac{\sum_{u' \in u, v' \in v}w_{u'v'}}{\sum_{u' \in u}w_{u'}} \mathrm{\ for\ } u',v' \in \chi^m \mathrm{\ and\ } u=s_u, v=s_v. \label{eq:lumpedtransitionprob}
\end{align}

There is a suit of inference methods for different classes of variable-length Markov chains, based on Kullback-Leibler distance,\cite{Machler:2012hd} cross validation,\cite{Xu:2016ex} and information-theoretic or Bayesian model selection.\cite{Jaaskinen:2014he,Xiong:2016co} However, with few exceptions,\cite{Xu:2016ex} they are, in our terminology, developed for small sets of physical nodes $\chi$, such as the four letters of DNA or 128 characters of ASCII text. Moreover, because our ultimate goal is statistically validated maps of network flows, we want to perform model selection on the maps rather than on the sparse Markov chain models themselves. Consequently, we cannot directly apply existing inference methods for constructing statistically validated maps of network flows based on thousands of physical nodes.

\begin{figure*}[thbp]
\centering
\includegraphics[width=0.8\textwidth]{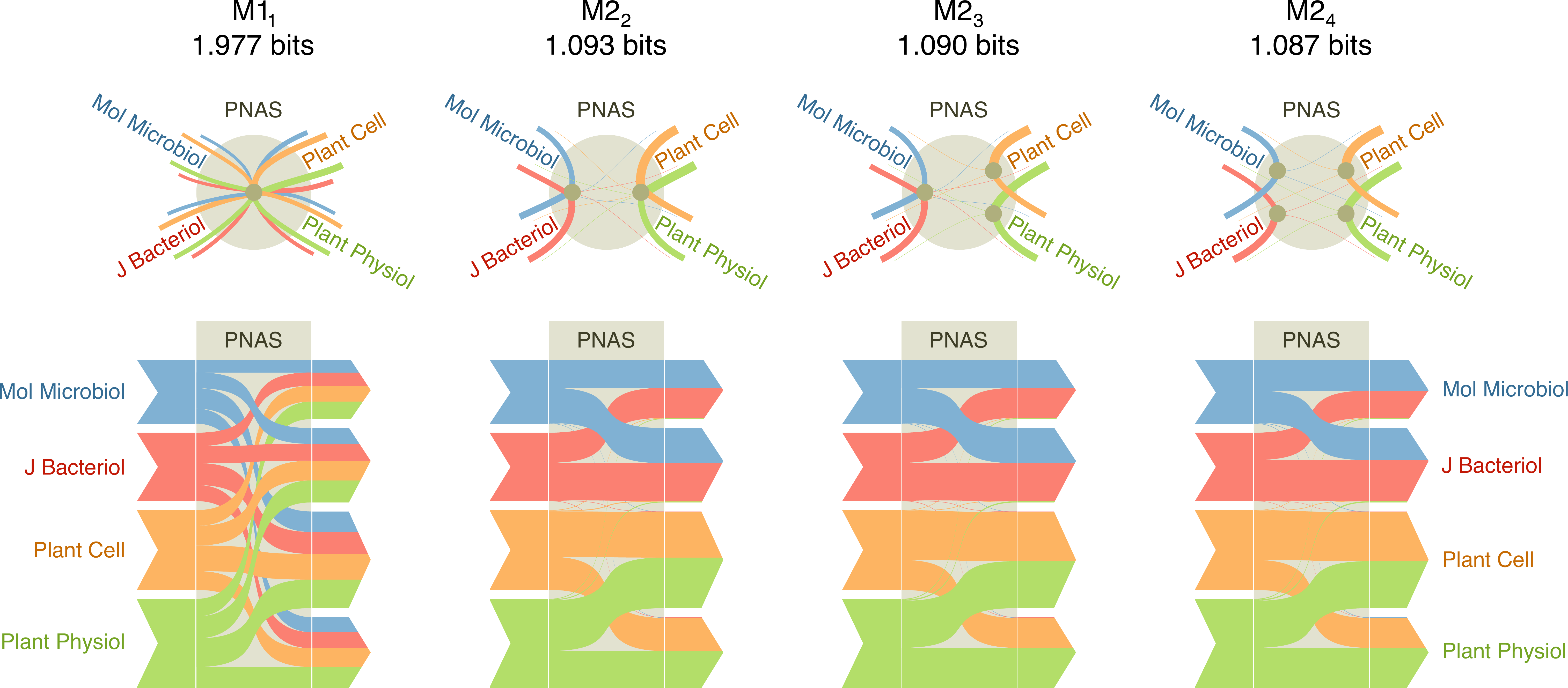}
\caption{The lumping algorithm generates sparse Markov chains with minimum information loss. Actual citation flows from four journals through PNAS modeled with one to four state nodes associated with the physical node. First-order Markov chain model with one state node to the very left, M1$_{1}$, and a fixed second-order Markov chain model with four state nodes to the very right, M2$_{4}$. The bit rates represent the entropy rates of the Markov chain models for the citation flows. The sparse Markov chain model with two state nodes, M2$_{2}$, efficiently captures that citation flows from the microbiology journals return to microbiology and that citation flows from plant science journals return to plant science. Despite that M2$_{2}$ only uses two state nodes, compared with the fixed second-order Markov chain model, M2$_{4}$, the information loss is less than one percent. When mapping all state nodes in all journals with the higher-order Markov chain models, the state nodes for the microbiology journals will be assigned to the same research field and the state nodes for the plant science journals will be assigned to another research field, thereby forming overlapping research fields in PNAS. \label{fig:lumping}}
\end{figure*}

Instead, we have developed a Markov chain lumping algorithm based on minimal information loss. In each step, illustrated in Fig.~\ref{fig:lumping} for an application to citation flows, the algorithm lumps the two state nodes $u$ and $v$ into lumped state node $\scaleleftright{\{}{u,v}{\}}$ that minimizes the entropy rate loss on the space of physical nodes. This is equivalent to minimizing the weighted sum of Kullback-Leibler distances between the non-lumped and lumped process for each lumped state node.
That is, with $P^{\scaleleftright{\{}{u}{\}},\scaleleftright{\{}{v}{\}}}_u$ for the transition probabilities of state node $u$ when state nodes $u$ and $v$ remain separate and $P^{\scaleleftright{\{}{u,v}{\}}}_{\scaleleftright{\{}{u,v}{\}}}$ when they are lumped, the algorithm identifies $u$ and $v$ such that
\begin{align}
\argmin_{u,v \in S}\, w_u D(P^{\scaleleftright{\{}{u}{\}},\scaleleftright{\{}{v}{\}}}_u \| P^{\scaleleftright{\{}{u,v}{\}}}_{\scaleleftright{\{}{u,v}{\}}} ) + w_v D(P^{\scaleleftright{\{}{u}{\}},\scaleleftright{\{}{v}{\}}}_v \| P^{\scaleleftright{\{}{u,v}{\}}}_{\scaleleftright{\{}{u,v}{\}}} ).
\end{align}
For memory efficiency, we apply this lumping algorithm to state nodes of each physical node separately, store the successive lumpings and associated entropy-rate increases with efficient data structures, and stop when there is only one state node left---the physical node itself with first-order transition probabilities.

To build the sparse Markov chain model for the entire system, we start with the first-order Markov chain model with $N$ state nodes, one for each physical node. Then we successively unlump the states, one by one, by choosing the unlumping in turn across all physical nodes that gives the largest decrease in entropy rate. Once we have reached a given number of state nodes, we build the network of links between all state nodes. In this way, we can build efficient sparse Markov chain models with strictly decreasing entropy rates for increasing number of state nodes. Ultimately we choose the one that according to model selection gives the best map of network flows given the multi-step pathway data. 

\subsection*{Mapping network flows modeled by sparse Markov chains.}
While sparse Markov chains provide efficient models of network flows, they do not  identify modules with long flow persistence times. Such flow modules have proved useful for identifying important functional structures in social and biological systems.\cite{rosvall2008maps,lancichinetti2009community,delvenne2010stability,rosvall2014memory} Among the different community-detection algorithms for identifying modules in network flows, particularly effective for sparse Markov chains is the so-called map equation. The information-theoretic map equation quantifies in bits how well a partition of nodes into possibly nested and overlapping modules can compress a description of flows on a network. Because compressing data is dual to finding regularities in the data,\cite{shannon1948mathematical} the modular description with maximum compression according to the map equation therefore is the one that best captures modular regularities in the flows on the network.

Importantly, the map equation can discriminate between physical nodes for describing the concrete objects that flow entities can visit and the abstract state nodes for representing the dynamics. This inherent feature follows because the map equation uses optimal codewords for relevant events, such as when flow entities enter and exit modules or visit physical nodes in a given module, but not for irrelevant events, such as to discriminate multiple state nodes of a physical node in a module.
In contrast, direct application of community-detection algorithms for first-order Markov chains to the extended state space of higher-order Markov chains confuse relevant and irrelevant states.\cite{salnikov2016using,Xu:2016ex}
Strictly speaking for the map equation, when two or more state nodes of the same physical node are assigned to the same module, the two state nodes share the same visit-frequency-derived codeword in the modular description. As illustrated with actual data in Fig.~\ref{fig:lumping}, when two plant scientists navigate scholarly literature from two different plant science journals to multidisciplinary \emph{PNAS}, they use a common codeword for \emph{PNAS} if their state nodes are assigned to the same module. In the same way, two microbiologists with different journal origins will use another common codeword when they visit \emph{PNAS} as long as their corresponding state nodes are assigned to the same module. As a result, the map equation's encoding that discriminates between physical nodes and state nodes is not only more natural, but also more efficient.

To discriminate relevant and irrelevant states when optimizing the map equation over possible module assignments, we have developed an updated version of the community-detection algorithm Infomap that can operate on state nodes associated with physical nodes.\cite{infomap} Given a sparse Markov chain model, or any model that can be represented with state nodes and physical nodes, Infomap searches for the state node assignments to hierarchically nested modules that give the shortest modular description of the data according to the map equation.

To identify the sparse Markov chain model that gives the statistically best map and  controls for under- and overfitting, we perform standard ten-fold cross-validation analysis.\cite{arlot2010survey} In general, we partition the data into ten subsets, combine nine of the sets into a training set and use the remaining set for validation. First, we build the sparse Markov chains for the training set with the lumping algorithm described above. Second, we apply the map equation's optimization algorithm Infomap\cite{infomap} and search for the possibly nested and overlapping modular description that minimizes the map equation. Third, we map the lumpings and mappings of the training set onto the validation set and measure the code length of this modular description. We repeat these three steps for the ten possible folds of training and test sets, and, starting with the sparsest first-order Markov chain model and moving in exponentially increasing steps to sparse Markov chain models with more states, we continue until the median validation code length in the third step reaches a minimum. In this way, we can identify the sparse Markov chain representation of the data for mapping the dynamics that is both statistically sound and computationally efficient. 

\subsection*{Mapping citation flows modeled by sparse Markov chains.}
To illustrate the advantages of mapping network flows modeled by sparse Markov chains, we use citation flows between scientific journals. Science is not only the systematic human quest for new knowledge about how things work in the universe, thanks to rich data it is also in itself an interesting model system for cultural and organizational evolution, and metascience, science of science, and bibliometrics are flourishing research fields.\cite{Evans721,wang2013quantifying,hicks2015bibliometrics} From researchers to research councils, stakeholders rely on good categorizations of scientific journals for efficient search and analysis of the humongous and ever-growing literature. However, in the most widely used journal categorization provided by the Thomson Reuters Web of Science, the subject categories are manually derived and curated based on unclear criteria.\cite{leydesdorff2009global,leydesdorff2013global} Ever since the 60's when de Solla Price postulated that journal citations can reveal the topography of science,\cite{de1965networks} researchers have suggested different methods to overcome the automation and transparency problems. However, these methods have either suffered from computational challenges that limits them to subsets of the scientific literature,\cite{pudovkin2002algorithmic,leydesdorff2006can} or they  exclude or assign journals only to single subject categories and thereby misclassify multidisciplinary journals.\cite{boyack2005mapping,rosvall2008maps,rafols2010science} Since multidisciplinary journals, such as Nature, PNAS, and Science, are top journals with a significant fraction of all citations, more accurate maps must capture the overlapping nature of scientific fields.

Mapping network flows modeled by sparse Markov chains applied to citation flows 
provides a transparent, automated, computationally efficient, and accurate method to identify scientific fields that overlap in multidisciplinary journals. We use citation data from the Thomson Reuters Web of Science, 1980--2013.\cite{jsrnote} Excluding proceedings, the data include almost one billion citations between more than 30 million articles published in about 20,000 journals. While different studies will need different selections of citation data for different maps, to avoid problems with old or scarce data in small journals, we focused on articles published between 2007 and 2012 in the 10,000 journals with highest impact factor. Furthermore, to be able to partition the data into disjoint subsets for cross-validation, we  derived all citation paths of length three that include an article published in 2009. By mapping these 160 million citation paths between 4.9 million articles to 39 million paths between journals, we obtained a second-order Markov chain model with 2.5 million state nodes and 39 million weighted links as input to the lumping algorithm.

\begin{figure}[thbp]
\centering
\includegraphics[width=\columnwidth]{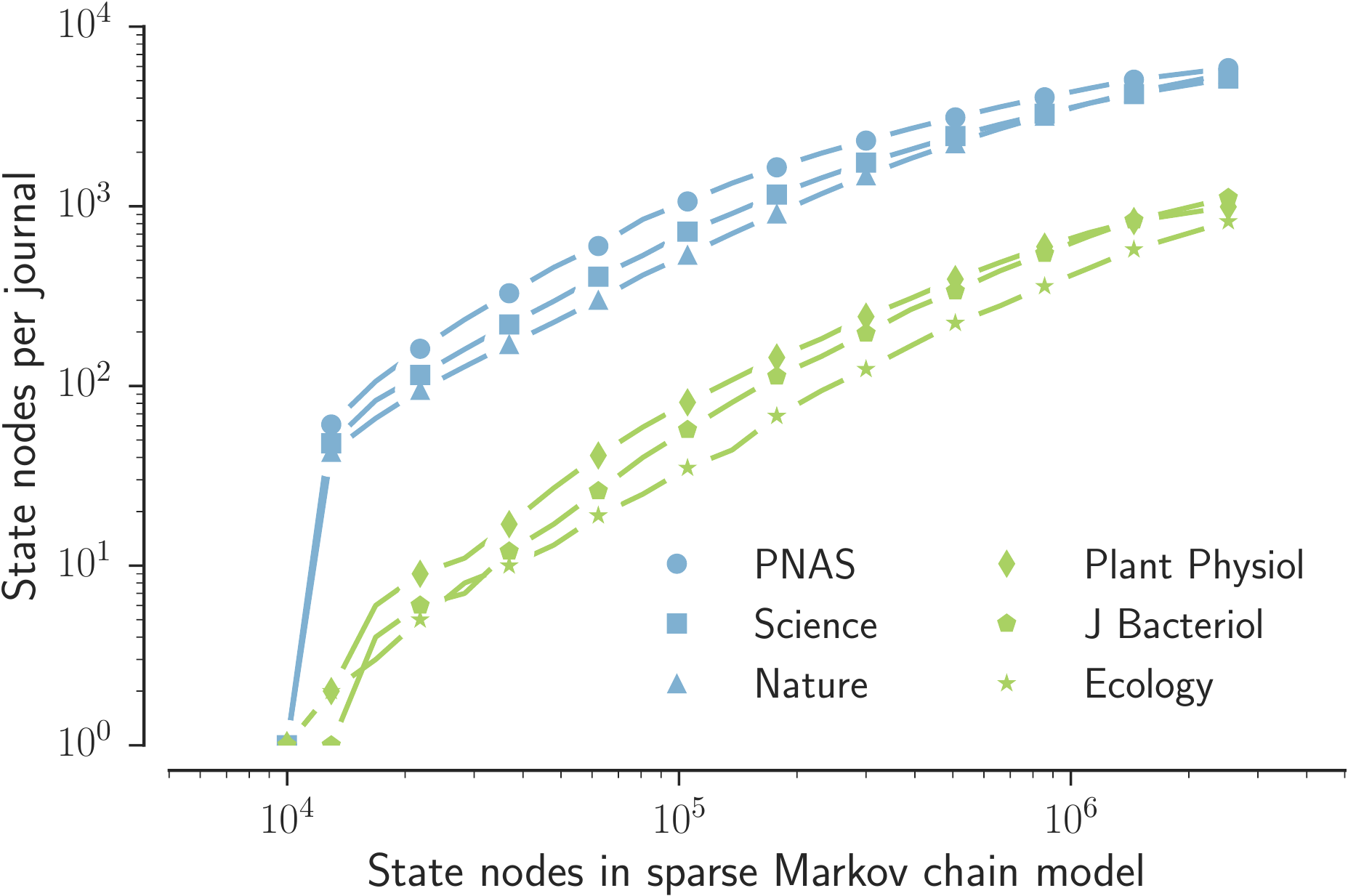}
\caption{The sparse Markov chain model allocates more state nodes to multidisciplinary journals than to specialized journals. In the first-order representation, each journal is trivially represented by a single state node. With more state nodes in the sparse Markov chain models, the multidisciplinary journals (black lines) use about ten times more state nodes than what specialized journals do (green lines). \label{statenodes_states}}
\end{figure}
We used the lumping algorithm to represent the dynamics with increasing number of state nodes in sparse Markov chain models from first- to second-order. Except for the trivial first-order case, the sparse Markov chain models represent multidisciplinary journals with about ten times more state nodes than specialized journals (Fig.~\ref{statenodes_states}). In this way, the sparse Markov chains efficiently take advantage of regularities in the data; the multidisciplinary journals are not only connected to more other journals, the out-link distributions of their state nodes also are more dissimilar from connections to diverse research fields.

\begin{figure}[thbp]
\centering
\includegraphics[width=\columnwidth]{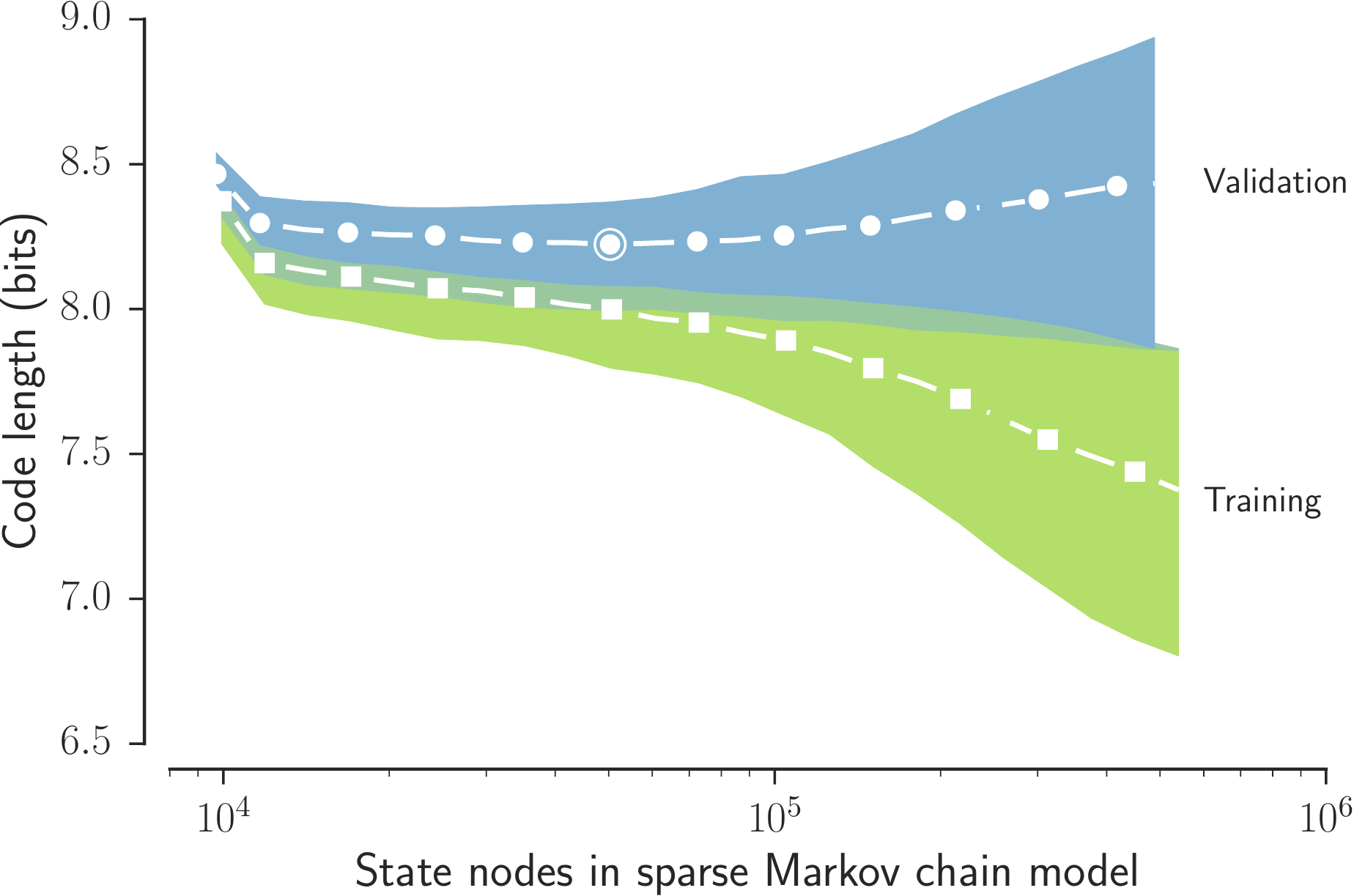}
\caption{A sparse Markov chain model with about 48,000 state nodes gives the best map of journal citation flows. Results based on ten-fold cross-validation. The lines represent the median code lengths and shaded bands contain all values. \label{fig:cross-validation}}
\end{figure}
We used the ten-fold cross-validation test to select the sparse Markov chains that give the statistically best citation flow map. Ideally, we want to divide the Markov chains into disjoint and independent subsets, and therefore used articles as the unit of sampling. For maximally independent subsets, however, the citation paths should be completely contained in article subsets sampled across all years, but this would drastically reduce the volume of data and make the remaining Markov chains insufficient for valid analysis. Instead, we obtained the ten disjoint subsets by, journal by journal for balanced sampling, randomly dividing the articles published in 2009 into one of ten equally sized subsets. Since each citation path contains an article in exactly one of these subsets, we obtained ten disjoint and, with respect to articles published in 2009, independent subsets of the Markov chains. Based on these subsets and the cross-validation schema described above, we looked for the number of state nodes that gives the shortest validation code length. The ten-fold cross-validation showed that a sparse Markov chain model with about 48,000 state nodes gives the best citation flow map (Fig.~\ref{fig:cross-validation}). 

\begin{figure}[thbp]
\centering
\includegraphics[width=1\columnwidth]{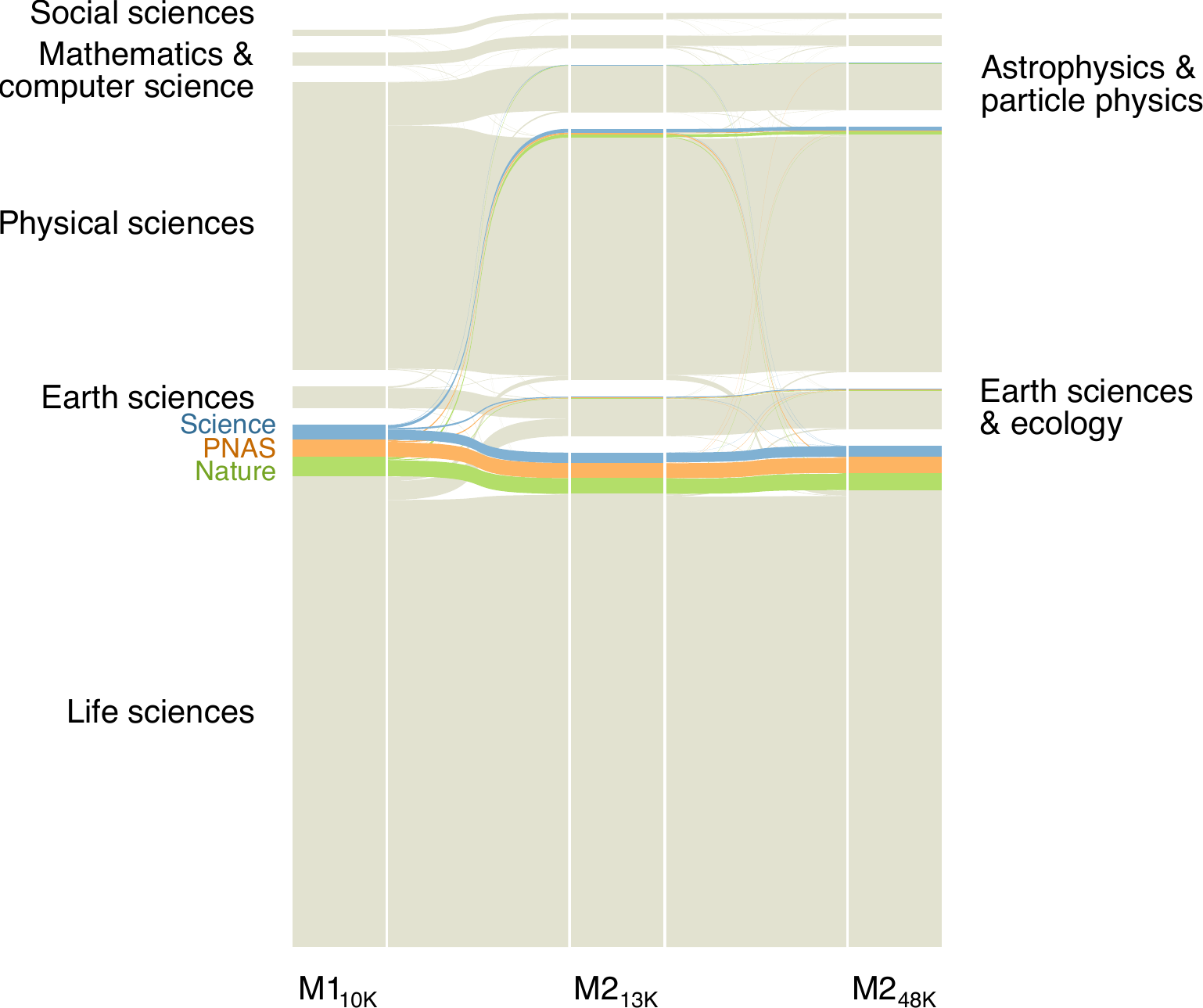}
\caption{Mapping citation flows modeled by sparse Markov chains gives research areas that overlap in multidisciplinary journals. Each block in a column of the alluvial diagram represents a research area at the highest clustering level,  and the height of the block reflects citation flows through the area. The different classifications of multidisciplinary journals Science, PNAS, and Nature are highlighted in blue, orange, and green, respectively. \label{fig:alluvial}}
\end{figure}
Citation flow maps based on sparse Markov chains gives robust research areas. We compiled all subsets of the Markov chains into a complete set, used the lumping algorithm to reduce the full second-order Markov chain model with more than 2.5 million state nodes and entropy rate 5.0 into a sparse Markov chain model M2$_{48\mathrm{K}}$ with about 48,000 state nodes and entropy rate 5.7, and used Infomap to search for the optimal hierarchically nested research fields. Counting research fields with more than one-hundredth percent of all citation flows, we identified six research areas at the highest level containing in total 226 research fields. Compared with the mapping of the standard first-order Markov chain model with 10,000 physical nodes, M1$_{10\mathrm{K}}$, astrophysics and particle physics emerges as a stand alone research area at the highest level (Fig.~\ref{fig:alluvial}). In fact, astrophysics and particle physics, as well as the integration of ecology into the earth sciences, appear already at the even more compact sparse Markov chain model M2$_{13\mathrm{K}}$ with entropy rate 6.0. The marginal difference on the observed code length in the cross-validation test reported in Fig.~\ref{fig:cross-validation} is in line with the robust research areas for the near optimal sparse Markov chains.

Citation flow maps based on the best sparse Markov chains give more well-defined research fields. We measured the module flow persistence, the fraction of citation flows that stays within the same research field in the next step, and compared the maps based on M2$_{48\mathrm{K}}$ and M1$_{10\mathrm{K}}$, as well as a journal classification provided by Thomson Reuters Web of Science.\cite{wang2015large} For multidisciplinary journals, we found that going from maps based on M1$_{10\mathrm{K}}$ to M2$_{48\mathrm{K}}$ improved the module flow persistence with 19 percent for Nature, 25 percent for PNAS, and 42 percent for Science. For research fields, the improvement was, for example, 3 percent for microbiology, 29 percent for molecular biology, and 44 percent for plant science. Overall, the module flow persistence across all research fields increased with 38 percent. When we compared with the classification by Web of Science with similar number of research fields, we identified journal classifications corresponding to 95 percent of all citation flows, and generously measured the module flow persistence as the fraction of citation flows that in the next step stays within any of the research fields a journal is assigned to. Nevertheless, the citation flow maps based on M2$_{48\mathrm{K}}$ showed 86 percent longer module flow persistence (82 percent compared with 44 percent module flow persistence). These results show that, compared with mapping first-order Markov chain and the established classification by Web of Science, mapping sparse Markov chains identifies superior research fields.

\begin{table*}[tbp]
\caption{Mapping citation flows modeled by sparse Markov chains reveals multidisciplinary journals. M1$_{\mathsf{10K}}$ for the standard first-order Markov chain model with 10,000 physical nodes and M2$_{\mathsf{48K}}$ for the sparse Markov chain model with 10,000 physical nodes and about 48,000 state nodes. Numbers represent the relative amount of citation flows that pass through the state nodes assigned to the respective research fields \label{table:journal_classifications}}
{\sffamily
\centering
\setlength{\tabcolsep}{2.9pt}
\rule{-0.7cm}{0cm}\scalebox{0.75}{\begin{tabular*}{1.18\hsize}{@{}rllllllllllllll@{}}\mytoprule\noalign{\smallskip}
{} & Science & & \rule{0.1cm}{0cm}& PNAS & &\rule{0.1cm}{0cm}& Nature && \rule{0.1cm}{0cm}& Plant Physiology && \rule{0.1cm}{0cm}& J Bacteriology & \\ \noalign{\smallskip}
& {M1$_{\mathsf{10K}}$} & {M2$_{\mathsf{50K}}$} && {M1$_{\mathsf{10K}}$} & {M2$_{\mathsf{50K}}$} && {M1$_{\mathsf{10K}}$} & {M2$_{\mathsf{50K}}$} && {M1$_{\mathsf{10K}}$} & {M2$_{\mathsf{50K}}$} && {M1$_{\mathsf{10K}}$} & {M2$_{\mathsf{50K}}$} \\ \noalign{\smallskip}
\mymidrule\noalign{\smallskip}
Molecular Biology 
& \colorrule{mygreen}{1.0cm}{0.15cm} 100 & \colorrule{myblue}{0.58cm}{0.15cm} 58 &
& \colorrule{mygreen}{1.0cm}{0.15cm} 100 & \colorrule{myblue}{0.72cm}{0.15cm} 72 &
& \colorrule{mygreen}{1.0cm}{0.15cm} 100 & \colorrule{myblue}{0.73cm}{0.15cm} 73 & 
&  -  &  -  & 
&  -  &  -  \\
Physics 
&  -  & \colorrule{myblue}{0.08cm}{0.15cm} 8 &
&  -  & \colorrule{myblue}{0.02cm}{0.15cm} 2 &
&  -  & \colorrule{myblue}{0.09cm}{0.15cm} 9 & 
&  -  &  -  &
&  -  &  -  \\
Nanotechnology 
&  -  & \colorrule{myblue}{0.1cm}{0.15cm} 10 &
&  -  & \colorrule{myblue}{0.02cm}{0.15cm} 2 &
&  -  & \colorrule{myblue}{0.05cm}{0.15cm} 5 & 
&  -  &  -  &
&  -  &  -  \\
Neuroscience 
&  -  & \colorrule{myblue}{0.04cm}{0.15cm} 4 &
&  -  & \colorrule{myblue}{0.08cm}{0.15cm} 8 &
&  -  &  -  & 
&  -  &  -  &
&  -  &  -  \\
Immunology 
&  -  & \colorrule{myblue}{0.03cm}{0.15cm} 3 &
&  -  & \colorrule{myblue}{0.05cm}{0.15cm} 5 &
&  -  & \colorrule{myblue}{0.06cm}{0.15cm} 6 & 
&  -  &  -  &
&  -  &  -  \\
Ecology 
&  -  & \colorrule{myblue}{0.02cm}{0.15cm} 2 &
&  -  & \colorrule{myblue}{0.03cm}{0.15cm} 3 &
&  -  & \colorrule{myblue}{0.01cm}{0.15cm} 1 & 
&  -  &  -  &
&  -  &  -  \\
Chemistry 
&  -  & \colorrule{myblue}{0.05cm}{0.15cm} 5 &
&  -  &  -  &
&  -  &  -  & 
&  -  &  -  &
&  -  &  -  \\
Plant Science 
&  -  & \colorrule{myblue}{0.01cm}{0.15cm} 1 &
&  -  & \colorrule{myblue}{0.02cm}{0.15cm} 2 &
&  -  & \colorrule{myblue}{0.01cm}{0.15cm} 1 & 
& \colorrule{mygreen}{1.0cm}{0.15cm} 100 & \colorrule{myblue}{1.0cm}{0.15cm} 100 &
&  -  &  -  \\
Astrophysics 
&  -  & \colorrule{myblue}{0.02cm}{0.15cm} 2 &
&  -  & \colorrule{myblue}{0.0cm}{0.15cm} 0 &
&  -  & \colorrule{myblue}{0.02cm}{0.15cm} 2 & 
&  -  &  -  &
&  -  &  -  \\
Microbiology 
&  -  &  -  &
&  -  & \colorrule{myblue}{0.02cm}{0.15cm} 2 &
&  -  &  -  & 
&  -  &  -  &
& \colorrule{mygreen}{1.0cm}{0.15cm} 100 & \colorrule{myblue}{1.0cm}{0.15cm} 100 \\
Others 
&  -  & \colorrule{myblue}{0.1cm}{0.15cm} 6 &
&  -  & \colorrule{myblue}{0.03cm}{0.15cm} 4 &
&  -  & \colorrule{myblue}{0.02cm}{0.15cm} 3 & 
&  -  &  -  &
&  -  &  -  \\
\mybottomrule
\end{tabular*}}
}
\end{table*}

Citation flow maps based on the best sparse Markov chain gives research fields that overlap in multidisciplinary journals. While maps based on citation flows modeled by M1$_{10\mathrm{K}}$ represent multidisciplinary journals only in single research fields, maps based on M2$_{48\mathrm{K}}$ represent them in multiple fields (Table~\ref{table:journal_classifications}). For example, Science is classified into ten research fields with at least one percent of Science's total citation flows, PNAS into nine fields, and Nature into seven fields. With respect to citation flows, molecular biology dominates in all three cases. On the other hand, specialized journals remain in single research fields as Table~\ref{table:journal_classifications} shows for J Bacteriology and Plant Physiology. Overall, better models of citation flows give better maps that capture the specialized or multidisciplinary nature of journals.

\begin{figure*}[tbp]
\centering
\includegraphics[width=\textwidth]{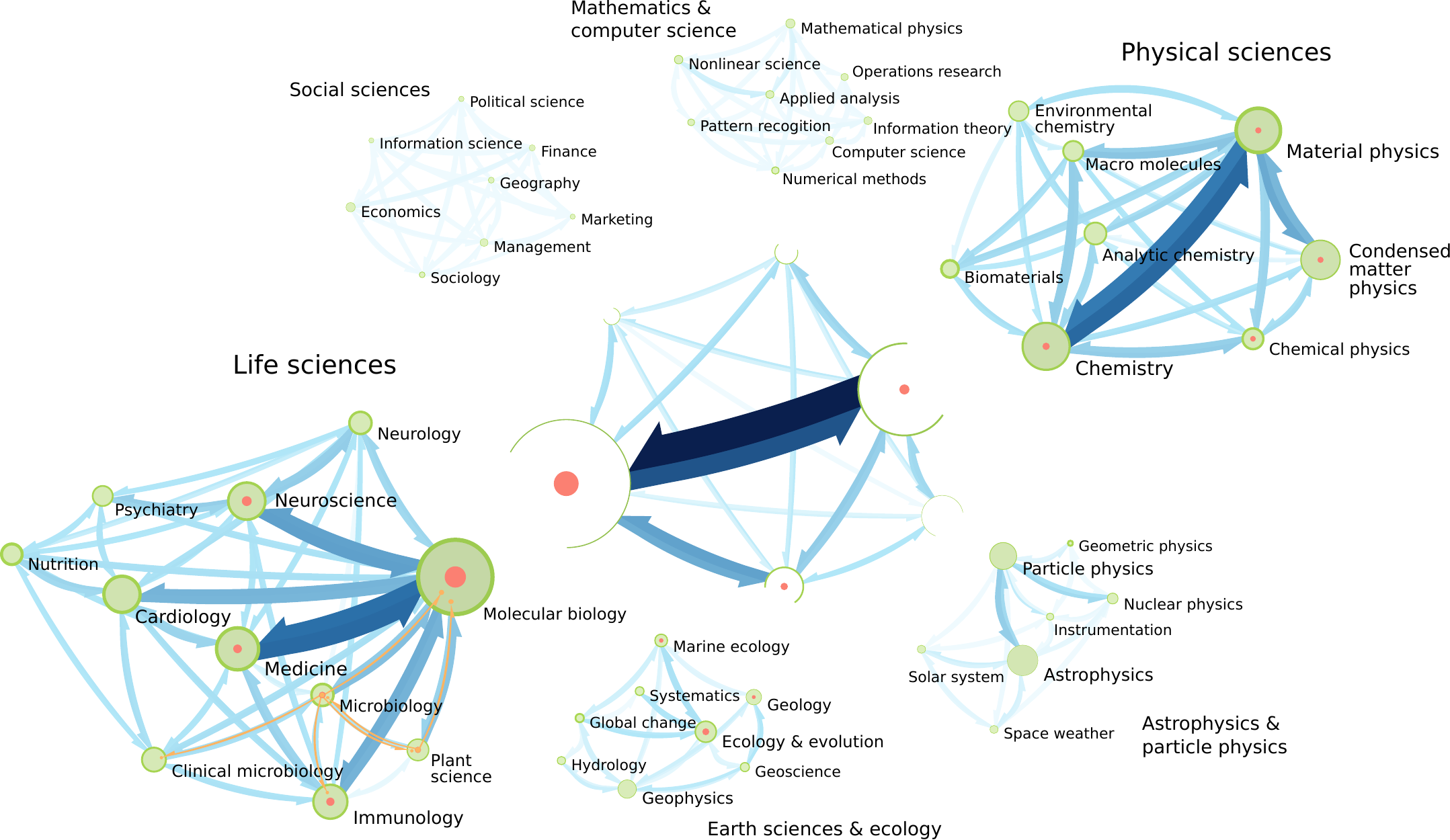}
\caption{Map of citation flows in science modeled by a sparse Markov chain model.\cite{statemap} The map highlights the most influential research fields in each research area (green circles), citation flows between fields (blue arrows), and how PNAS is represented in multiple research fields (red circles). Orange circles indicate where citation flows through PNAS represented in microbiology and plant science move next, respectively. Circle sizes are proportional to the citation flows and the inner circles represent module flow persistence. The map builds on sparse Markov chain model M2$_{13\mathrm{K}}$ based on citation flows through 4.9 million articles published 2007--2012 in the top 10,000 journals of Web of Science. \label{fig:map_of_science}}
\end{figure*}
Sparse Markov chains give better maps with respect to both citation flows and journal classifications. In Fig.~\ref{fig:map_of_science}, we represent the citation flows in science as a visual map. It is a static version of an interactive map available online, and for responsiveness we therefore base it on the more compact sparse Markov chains in M2$_{13\mathrm{K}}$.\cite{statemap} As Fig.~\ref{fig:alluvial} indicates, this map is similar to the optimal map based on M2$_{48\mathrm{K}}$. In any case, representing multidisciplinary journals in multiple fields and modeling the citation flows with sparse Markov chains together improve the module flow persistence. For example, the citation flows through PNAS represented in plant science (red circle almost covered by orange circle) mostly persist in plant science (orange circle), with limited citation flows mostly to microbiology and molecular biology. Similarly, the citation flows through PNAS represented in microbiology mostly persist in microbiology, with limited citation flows to neighboring research fields. In contrast, in a first-order Markov chain model with PNAS represented only in molecular biology, citation flows from microbiology or plant science would cross research boundaries when moving through PNAS in molecular biology. This example illustrates how sparse Markov chains with multiple state nodes per physical node enable context dependent flows and maps with overlapping modules, which in turn are essential for efficient modular descriptions of network flows with memory.

\section*{Conclusions}
We have designed maps of network flows modeled by sparse Markov chains with several advantages: The sparse Markov chains derived from an efficient lumping algorithm compactly represent memory in network flows and the cross-validated maps reveal hierarchically nested and overlapping communities. Furthermore, the method applies to multi-step pathways of any length. For illustration, based on citation flows through 4.9 million scientific articles in 160 million citation pathways, we classified 10,000 journals into hierarchically nested research fields that overlap in multidisciplinary journals. With small computational overhead, the quality of the journal classification, measured as the fraction of citation flows that stays within the same research field in the next step, increased from 59 percent for a conventional mapping based on first-order Markov chains to 82 percent. Compared with an established classification by Web of Science, which has a module flow persistence of 44 percent, the data-derived journal classification should form better units of analysis for building recommendation systems, comparing impact across research fields, or measuring interdisciplinarity in science of science studies. We anticipate that other systems will see similar benefits from efficient maps of sparse Markov chains.

\section*{Acknowledgements}
M.R.\ was supported by the Swedish Research Council grant 2012-3729. The computations were performed on resources provided by the Swedish National Infrastructure for Computing (SNIC) at High Performance Computing Center North (HPC2N). We thank Qi Wang and Ludo Waltman for providing classification data from Thomson Reuters Web of Science.



\end{document}